\documentclass[11pt]{article}

\usepackage[dvips]{graphicx}
\usepackage{latexsym}
\usepackage{amsfonts}

\newcommand{\be}{\begin{equation}}
\newcommand{\ee}{\end{equation}}
\newcommand{\ra}{\rightarrow}
\newcommand{\wt}{\widetilde}
\newcommand{\vp}{\varphi}
\newcommand{\wi}{\omega_{\infty}}
\newcommand{\si}{\Phi_{\infty}}
\newcommand{\ex}{\times 10}
\newcommand{\integ}{\int_{0}^{\infty}}
\newcommand{\smfrac}[2]{{\textstyle{#1\over#2}}}
\def\half{\smfrac{1}{2}}

\begin{document}
\title{Spontaneous Scalarization and Boson Stars}

\author{A. W. Whinnett\thanks{Email: A.W.Whinnett@qmw.ac.uk}\\
        School of Mathematical Sciences,\\ Queen Mary and Westfield
College,\\ University of London,\\ United Kingdom}

\maketitle

\begin{abstract}

We study spontaneous scalarization in Scalar-Tensor
boson stars. We find that
scalarization does not occur in stars whose bosons have no
self-interaction. We introduce a quartic self-interaction term into
the boson Lagrangian and show that when this term is large,
scalarization does occur. Strong self-interaction leads to a large
value of the compactness (or sensitivity)
of the boson star, a necessary condition for
scalarization to occur, and we derive an analytical expression for
computing the sensitivity of a boson star in Brans-Dicke theory from
its mass and particle number. Next we comment on how one can use
the sensitivity of a star in any Scalar-Tensor theory to determine how
its mass changes when it undergoes gravitational evolution. Finally,
in the Appendix, we derive the most general form of the boson
wavefunction that minimises the energy of the star when the bosons
carry a $U(1)$ charge.

\end{abstract}


\section{Introduction}\label{INT}

Scalar Tensor (ST) theories of gravity are the most natural
generalisations of General Relativity (GR) and describe gravity as
being mediated by both a metric field $g_{ab}$ and a scalar field
$\Phi$. In the simplest ST theories, Newton's constant $G$ is replaced
by a field $G\Phi^{-1}$ at the level of the action, while
$\Phi$ obeys a wave equation and is coupled to the curvature via a
function $\omega(\Phi)$. The strength of this coupling decreases as
$\omega$ increases and the special case of constant $\omega$ gives
the Brans Dicke (BD) theory. One recovers GR by taking the limit
$\omega\ra\infty$, $\Phi\ra\Phi_{GR}$, where $\Phi_{GR}$ is
constant. By choosing appropriate units, one can set $\Phi_{GR}=1$.
In the original (Jordan frame)
formulation of these theories, the ordinary matter is universally
coupled to the curvature, so these theories are metric theories.
As is well known, a conformal transformation may be used to express the
theory in a new set of variables $g_{ab}^{*}=\Phi g_{ab}$,
$\vp=\vp(\Phi)$, where $g_{ab}^{*}$ is the Einstein frame metric and
$\vp$ is a new scalar field that plays the role of a matter source and
also couples the matter to $g_{ab}^{*}$. The Einstein frame
representation is often more convenient mathematically than the Jordan
frame (for example, the Cauchy problem is well posed in the
Einstein frame \cite{DEF1}) although the Jordan frame is often
considered to be the physical one.

At present, the primary motive for studying ST gravity comes from
supergravity and superstring theories whose low energy limits always
include one or more scalar fields (dilaton fields) that play a similar
role to the $\Phi$ field of ST gravity. However, the 
coupling between the non-dilaton terms and the curvature in string
actions is non-universal so that the
correspondence between the low energy limit of supersymmetric theories
and ST gravity is only approximate. Nevertheless, this approximation
is fairly accurate since, to date, highly sensitive experiments have
detected no violation of the principle of
universal coupling \cite{Will2}.

In general, the results of Solar System experiments place the
strongest constraints on the viability of most theories of
gravity. For metric theories, these experiments are interpreted within
the framework of the parameterised post-Newtonian (PPN) formalism
which describes the gravitational field of the Solar System as an
expansion about flat spacetime in powers of the Newtonian 
gravitational potential $U$ near to the surface of the Sun. 
The coefficients appearing in the expansion up to the order $U^{4}$
terms, the so called first post Newtonian (1PN) order,
are the 10 PPN parameters. In ST gravity only two of
these are non-zero: $\gamma$ and $\beta$, while $\Phi$ is everywhere
close to some cosmologically determined boundary value $\Phi_{B}$. 
In terms of the coupling function $\omega$ the two PPN parameters 
of ST theory are given by
\be\label{PPNpars}
   \gamma=\frac{1+\omega_{B}}{2+\omega_{B}},\;\;\;\;
   \beta=1+(2\omega_{B}+3)^{-2}(2\omega_{B}+4)^{-1}\frac{d\omega}{d\Phi},
\ee
where $\omega_{B}=\omega(\Phi_{B})$ and $d\omega/d\Phi$ is to be evaluated at
$\Phi=\Phi_{B}$. 
The observational constraints on $\gamma$ and $\beta$ are \cite{Will2}
\be\label{limits}
   |\gamma-1|\leq 0.0003, \;\;\;\;
   |\beta-1|\leq 0.002.
\ee
For BD theory, the first of these inequalities implies that $\omega\ge
3300$ and all of the predictions of BD theory differ from those of GR to
within a relative error of $\sim 1/\omega$.
For other ST theories, the observational data only places
limits on the behaviour of $\omega(\Phi)$ in the slow motion, weak
field limit.

A framework for analysing the motion of compact relativistic bodies in metric
theories of gravity, the modified EIH formalism, has been
developed (see Eardley \cite{Eardley} and Will \cite{Will} for
details). The formalism treats each body as a point mass moving in the
gravitational field of its companion bodies. Each body has an inertial
mass which governs its response to the ambient gravitational field via
quasi-Newtonian equations of motion. The role of non-metric
gravitational fields (such as the $\Phi$ field in ST gravity) is to
modify the inertial mass. The modified EIH formalism forms the basis
of the post Keplerian formalism, which gives a parameterised
description of the motion of binary pulsar systems and is used to interpret
observational data gathered from these systems. 
These data must be used to constrain or rule out
theories of gravity on a case by case basis (in contrast with solar
system observations which can place limits on entire classes of
theories by limiting the values of the PPN parameters). 

Since the coupling function $\omega$ is completely arbitrary,
there are an infinite number of viable ST theories. This leaves open
the possibility
that gravity is described by a ST theory whose predictions are
arbitrarily close to those of GR at the current epoch and in the weak
field limit, but differ considerably in strong field situations or at
earlier cosmological epochs. Thus astrophysical objects with strong
internal gravitational fields may still exhibit behaviour that differs
greatly from that predicted by GR.

One strong field effect is spontaneous scalarization, recently
discovered in neutron stars by Damour \& Esposito-Farese
\cite{DamourEF,DEF2}. They found that, for ST theories in which the
coupling function obeys the inequality
\be\label{betalimit}
   {\wt\beta}:=\frac{2\Phi_{B}}{(2\omega+3)^{2}}\left.
   \frac{d\omega}{d\Phi}\right|_{\Phi=\Phi_{B}}< -4
\ee
and satisfies the constraints of eqn (\ref{limits}), the Einstein
frame scalar field inside a neutron star rapidly becomes inhomogeneous once
the star's mass increases above some critical value. For a star whose
mass is below this value, $\vp$ is nearly constant throughout the
star (a state which minimises the star's energy), while for higher
mass stars, the energy is minimised when $\vp$ has a large spatial
variation. These effects become more pronounced
in the limit $\Phi_{B}\ra\Phi_{GR}$,
$\omega_{B}\ra\infty$. The coupling function they chose was of the
form $2\omega+3\sim 1/(\log\Phi)$ which diverges as $\Phi\ra\Phi_{GR}=1$
and gives weak field neutron stars that are indistinguishable from
their GR counterparts. A similar study was performed by
Salgado, Sudarsky \& Nucamendi \cite{SalSud}, who considered neutron
stars in a theory with a non-linearly coupled scalar field $\chi$
which is equivalent to standard ST theory with
$\omega\sim\Phi/(\Phi-1)$, which also diverges as $\Phi\ra\Phi_{GR}$.
Independently of Damour \&
Esposito-Farese, a study of strong field effects in ST neutron stars
was carried out by Zaglauer \cite{Zag}. However, the formalism 
developed in \cite{Zag} involved an artificial
definition of the sensitivity (defined below) of a neutron star,
leading to the apparent occurrence of strong ST effects for choices of
$\omega$ and $\Phi_{B}$ that the study in \cite{DamourEF} showed 
should not occur.

It turns out that when spontaneous scalarization effects are present
in neutron stars,
the modified EIH formalism breaks down. An alternative, more
sophisticated formalism, adapted exclusively to tensor-multi-scalar
theories of gravity, has recently been devised by Damour \&
Esposito-Farese \cite{DEF1}. Their formalism, which is based on an
expansion of an effective coupling strength $\alpha$ 
between the $\vp$ field and the
non-scalar matter about its weak field value, is well suited to the study
of the scalarization phenomenon.

The purpose of the present paper is twofold. First we discuss some
aspects of the
modified EIH formalism as it applies to ST gravity after introducing
appropriate definitions of the inertial mass and sensitivity of a
star. We then numerically compute the sensitivity of boson stars for
several choices of $\omega$. Secondly, we demonstrate the existence of
spontaneous scalarization in boson stars and compare the behaviour of
the sensitivity and the coupling parameter $\alpha$ with their
behaviour in neutron stars.

\section{Field Equations and Scalar Field Boundary Values}\label{FORM}

The Jordan frame action for scalar-tensor gravity 
universally coupled to one or more
matter fields $\mu$ is given by
\be\label{JFaction}
   I=\int d^{4}x\sqrt{-g}\left[\frac{1}{16\pi}\left(\Phi R
   -\frac{\omega(\Phi)}{\Phi}g^{ab}\partial_{a}\Phi\partial_{b}\Phi\right)
   +L_{m}(g^{ab},\mu)\right]
\ee
where $g_{ab}$ is the Jordan frame metric, $R$ is the Ricci scalar curvature 
formed from $g_{ab}$ and we use units in which $G=c=1$. The
generalised Einstein equations in this frame are 
found by varying $I$ with respect to $g^{ab}$ and are given by
\be\label{JFfeqns}
   G_{ab}=\frac{1}{\Phi}(\nabla_{a}\nabla_{b}\Phi 
   -g_{ab}g^{cd}\nabla_{c}\nabla_{d}\Phi)
   +\frac{\omega}{\Phi^{2}}
   (\partial_{a}\Phi\partial_{b}\Phi-\half g_{ab}g^{cd}
    \partial_{c}\Phi\partial_{d}\Phi)
   +\frac{8\pi}{\Phi}T_{ab}
\ee
where
\be\label{JFTdef}
   T_{ab}:=-\frac{2}{\sqrt{-g}}
   \frac{\delta(\sqrt{-g}L_{m})}{\delta g^{ab}}
\ee
is the matter energy-momentum tensor,
$G_{ab}$ is the Einstein tensor formed from
$g_{ab}$ and $\nabla_{a}$ is the covariant derivative
compatible with $g_{ab}$. The scalar and matter fields obey the equations
\be\label{JFothereqns}
   g^{ab}\nabla_{a}\nabla_{b}\Phi
   =\frac{1}{2\omega+3}\left(8\pi T-\frac{d\omega}{d\Phi}
   g^{ab}\partial_{a}\Phi\partial_{b}\Phi\right),\;\;\;\;
   \nabla_{a}T^{ab}=0
\ee
where $T:=g^{ab}T_{ab}$.

To rewrite the field equations in the Einstein frame,
we make the field redefinitions
\be\label{conftrans}
   g^{*}_{ab}=A^{-2}g_{ab}, \;\;\;\; g_{*}^{ab}=A^{2}g^{ab},\;\;\;\;
   d\vp=\frac{\sqrt{2\omega+3}}{2\Phi}\;d\Phi,
\ee
where 
\be
   A^{2}(\vp):=\Phi^{-1}
\ee
is found by integrating the third of eqns (\ref{conftrans}) and
solving for $\Phi$. 
In the Einstein frame, the generalised Einstein equations are
\be\label{EFfeqns}
   G^{*}_{ab}=2\partial_{a}\vp\partial_{b}\vp-g^{*}_{ab}g_{*}^{cd}
   \partial_{c}\vp\partial_{d}\vp+8\pi T^{*}_{ab}
\ee
where 
\be\label{Tdef}
   T^{*}_{ab}:=A^{2}T_{ab},
\ee
while the matter and scalar field equations are
\be\label{EFothereqns}
   g_{*}^{ab}\nabla^{*}_{a}\nabla^{*}_{b}\vp
   =-\frac{4\pi}{A}\;\frac{dA}{d\vp}\;T^{*},\;\;\;\;\nabla^{*}_{a}T_{*}^{ab}=
   -\frac{16\pi}{A}\frac{dA}{d\vp}\partial_{a}\vp\;T_{*}^{ab},
\ee
where $T^{*}:=g^{ab}T^{*}_{ab}$ and $\nabla^{*}_{a}$ is the covariant 
derivative compatible with $g^{*}_{ab}$.

We analyse the behaviour of boson stars for three particular coupling functions
$\omega$. The first is the BD coupling, where $\omega$ is constant. 
The second is the exponential coupling law chosen in
\cite{DamourEF}, which in the Einstein frame is
\be\label{EFexp}
   A(\vp)=\exp(-k\vp^{2}),
\ee
where $k$ is a constant. This is equivalent to the Jordan frame coupling
\be\label{JFexp}
   2\omega+3=\frac{1}{2k\log\Phi}.
\ee
Our third choice is equivalent to that made by 
Salgado {\it et al}, who consider a theory in which the
Lagrangian is given by
\be\label{nonlinL}
   L=\frac{1}{16\pi}\left[(1+\xi\chi^{2})R
   -\frac{1}{2}g^{ab}\partial_{a}\chi\partial_{b}\chi\right]+L_{m},
\ee
where $\xi$ is a constant and $\chi$ a non-linearly coupled scalar
field
which is constrained to have the value $\chi\geq 0$.
Making the field redefinition $1+\xi\chi^{2}=\Phi$, one finds that the
theory derived from eqn (\ref{nonlinL}) is the standard ST theory
outlined above with the coupling function
\be\label{nlcoupling}
   \omega=\frac{\Phi}{8\xi(\Phi-1)},
\ee
where $\Phi\geq 1$. It is more natural to choose
$\Phi$ instead of $\chi$ as the physical scalar gravitational field
since, when studying the motion of test bodies about a gravitating
central source, it is $\Phi$ that appears explicitly in the equations
of motion for the test bodies.

We consider asymptotically flat solutions in which $\Phi$ tends
towards some limiting value $\si$ at spacelike infinity. We
assume that a star embedded in a cosmological background can be
approximated by an asymptotically flat solution and we identify
$\si$ with $\Phi_{B}$ at the present epoch. Then
equation (\ref{limits}) places restrictions on the maximum allowed
value of $\si$ for any solution at the current epoch. 
Using eqn (\ref{PPNpars}),
one finds that for the coupling given by eqn
(\ref{JFexp}), the maximum allowed boundary value is given by the
lower of the two values
\be\label{explimits}
   \Phi^{max}_{\infty}=1+\frac{0.0003}{4k},\;\;\;\;
   \Phi^{max}_{\infty}=1+\frac{0.002}{2k^{2}}.
\ee
Similarly, for the coupling function given by eqn (\ref{nlcoupling}),
the limiting value of $\si$ is given by the lower of the two
values
\be\label{nllimits}
   \Phi^{max}_{\infty}=1+\frac{0.0003}{8\xi},\;\;\;\;
   \Phi^{max}_{\infty}=1+\frac{0.002}{8\xi^{2}}.
\ee

\section{Mass and Scalar Charge}\label{MSC}

To study the mass of a boson star, or any other stellar object, we may
consider it either as a central source of gravitational field and explore the
surrounding spacetime with test particles of differing compositions,
or consider the star to be an extended, self-gravitating test body and
examine its motion in the gravitational field of a second massive
body. Either situation allows one to define several related quantities
based upon the mass of the star and the behaviour of $\Phi$ in the far
field region. The former situation was assumed by Salgado {\it et
al}.\   \cite{SalSud}, who
examined the mass and scalar charge (defined below) of neutron stars
considered as central sources. The latter situation is described by
the modified EIH formalism and the formalism devised by Damour and
Esposito-Farese and is considered by them in
\cite{DamourEF,DEF2}. We shall briefly summarise
definitions of mass and scalar charge for both situations, as applied
to any compact object (such as a boson star or neutron star), and
outline part of the modified EIH formalism. 
We specialise to systems consisting of spherically
symmetric bodies each of whose line elements, when described in the
Jordan frame, are given by  
\be\label{linel}
   ds^{2}=-B(r)dt^{2}+\frac{1}{1-2m(r)/r}dr^{2}+r^{2}dS^{2}
\ee
where $dS^{2}$ is the line element of the unit 2-sphere and $m(r)$
is the usual generalised Schwarzschild mass. Each body is modelled
as a static, asymptotically flat object.

Consider first the motion of test particles about a central star.
We define the Jordan frame scalar charge of the star by
\be\label{JFscharge}
   Q_{S}:=\lim_{r\to\infty}\left[r^{2}\frac{d\Phi}{dr}\right].
\ee
Taking the ADM mass $M_{ADM}:=lim_{r\to\infty}m(r)$ together with the
scalar charge, we define the tensor mass
\be\label{defMT}
   M_{T}:=M_{ADM}-\frac{Q_{S}}{2\si}
\ee
where $\si:=\lim_{r\to\infty}\Phi$ is the asymptotic  
value of the gravitational scalar field. 
$M_{T}$ measures the total energy of the star
and is also the active gravitational mass measured by an orbiting test
particle for which $\Phi$ is locally homogeneous (such as a small
black hole). An orbiting test particle with negligible self-energy moves on  
geodesics of $g_{ab}$ and measures the Keplerian mass
\be
   M_{K}:=M_{ADM}-\frac{Q_{S}}{\si}.
\ee
This is different from $M_{T}$ since the Strong Equivalence Principle 
(SEP, defined in \cite{Will}) is violated in ST gravity. 

From the expressions for $M_{K}$ and $M_{T}$ we have
\be
   Q_{S}=2\si(M_{T}-M_{K}).
\ee
A locally freely falling Cavendish experiment far from the star
measures a gravitational coupling strength
\be
   {\wt G}=\Phi^{-1}_{\infty}\frac{2\wi+4}{2\wi+3}
\ee 
where $\wi=\omega(\si)$. Given a particular form of
$\omega$, one can then calculate $\si$. Hence by combining the
above two equations, we find that $Q_{S}$ is measurable in principle.

Salgado {\it et al}. define and use the scalar charge
\be\label{nlscharge}
   {\wt Q}_{S}=\lim_{r\to\infty}
   \left[r^{2}\frac{d\chi}{dr}\right].
\ee
associated with the field $\chi$ that appears in the 
Lagrangian (\ref{nonlinL}). 
This quantity is related to the charge $Q_{S}$ associated with $\Phi$
by
\be
   Q_{S}=2\xi\chi{\wt Q}_{S}=2{\wt Q}_{S}\sqrt{\xi(\Phi_{\infty}-1)}.
\ee
From the above definitions it appears more natural to use $Q_{S}$ as
the physical scalar charge of the star instead of ${\wt Q}_{S}$.

Both $M_{T}$ and $Q_{S}$ may be related to corresponding
Einstein frame quantities as follows. In the Einstein frame, the
Schwarzschild radial coordinate $r^{*}$ is given by
$r^{*}=\sqrt{\Phi_{\infty}}\,r$. Then, defining the 
Einstein frame scalar charge
$Q^{*}_{S}$ by
\be\label{EFscharge}
   Q_{S}^{*}:=\lim_{r^{*}\to\infty}\left(r^{*2}\frac{d\vp}{dr^{*}}\right)
\ee
and using the third of eqns (\ref{conftrans}) gives
\be\label{QSrelate}
   Q_{S}^{*}=\frac{\sqrt{2\wi+3}}{2\sqrt{\si}}Q_{S}.
\ee
One can also show that $M_{T}$ is related to the Einstein frame ADM
mass $M^{*}_{ADM}$ by
\be\label{Mrelate}
   M^{*}_{ADM}=\sqrt{\Phi_{\infty}}\,M_{T}.
\ee 

Consider now the motion of a spherical star, with some conserved
particle number $N$, in the gravitational field
of a companion body. We assume that the separation between the two
bodies is much greater that either of their radii, and we denote the
inter body value of the scalar field by $\si$. The scalar
field $\Phi_{L}$ in the interior of the star must be matched smoothly
to $\si$ and its value will affect the structure of the
star. In particular, its inertial mass $M_{I}$ and hence its motion in
the gravitational field of the companion body will be sensitive to
changes in the value of $\si$. Expanding $\Phi_{L}$ about
$\si$ as $\Phi_{L}=\si+\delta\Phi$, we write
\be\label{MIexpand}
   M_{I}=M_{{I}_{\infty}}+\frac{\partial
   M_{I}}{\partial\Phi}\delta\Phi
   +{\cal O}(\delta\Phi^{2}),
\ee
where $M_{{I}_{\infty}}$ is the value of $M_{I}$ when
$\Phi_{L}=\si$, or
\be
   M_{I}=M_{{I}_{\infty}}\left[1+s\frac{\delta\Phi}{\si}\right]
   +{\cal O}(\delta\Phi^{2}),
\ee
where $s$ is the first sensitivity defined by
\be\label{sdef}
   s:=\frac{\partial(\log M_{I})}{\partial(\log\Phi)}.
\ee
In eqns (\ref{MIexpand}) and (\ref{sdef}), the derivatives are evaluated
at $\Phi=\si$ for fixed $N$. Higher order sensitivities may be
defined by taking successively higher order derivatives of
$M_{I}$, to obtain the expansion for $M_{I}$ to higher powers of
$\delta\Phi$. $s$ measures the response of a star's inertial mass to
changes in the value of $\si$ and it also gives a useful measure of
the compactness of the star \cite{DEF1}.

Here we define $M_{I}$ to be 
\be\label{MIdef}
   M_{I}:=\si M_{T}.
\ee
This differs from the definition given by Will \cite{Will} by a factor of
$(2\wi+3)/(2\wi+4)$ and is chosen so that our definition corresponds
with the Einstein frame quantity defined by Damour \& Esposito-Farese
in \cite{DEF1}. The difference
in definitions of $M_{I}$ does not affect the BD analysis given in
\cite{Will} since, for constant $\omega$, this factor cancels in the
definition of $s$ given there. For more general ST theories,
our definition if $M_{I}$ leads to a self consistent set of
equations of motion that, to 1PK order, are identical
to the BD equations derived in
\cite{Will} (except for the different definition of $M_{I}$)
and obviates the need to introduce the modification of $s$
given in \cite{Zag}.

These definitions form the basis of the modified EIH formalism, examined in
detail for more general theories of gravity in \cite{Will}, and
$s$ and its derivatives play a fundamental role in determining the motion
of a body in the gravitational field of other bodies. Formally, to
derive the equations of motion governing such a system, one 
first determines the inter body
gravitational fields from the field equations where the $M_{I}$ appear
as delta function terms in $T_{ab}$. One then substitutes the solution
into a quasi-Newtonian particle Lagrangian, taking into account
changes in the $M_{I}$ due to the relative velocities of the bodies,
to find the equations of motion for each body. These equations
explicitly involve the sensitivities of the bodies, 
which determine how the motion deviates from the Newtonian
prediction, and are written as an expansion in powers of $v/c$ about
the Newtonian solution, where $v$ is a characteristic velocity of the
system. (One cannot expand in powers of the Newtonian potential 
of the bodies since for strongly bound objects, $U$ is of the order of
1). The expansion is normally taken to order $v^{4}/c^{4}$, the first
post Keplerian (1PK) order, and in the limit that $U$ for each body
becomes small, the equations of motion reduce to the 1PN
equations. The post Keplerian equations then are a parameterised
description of the modified EIH equations of motion, adapted so that
the system is described directly in terms of observable parameters.

For a system composed of two bodies it turns out that only the first
sensitivity $s$ appears in the equations of motion to 1PK order. 
For example, as shown in detail in \cite{Will},
given a star of inertial mass $M_{I}^{A}$ and sensitivity $s_{A}$ with
a companion  of mass $M_{I}^{B}$ and sensitivity $s_{B}$ at a distance
$R_{AB}$, the quasi-Newtonian
gravitational force $F$ between the two bodies has a magnitude
$F=G_{AB}M_{I}^{A}M_{I}^{B}/R_{AB}^{2}$ where the effective
gravitational coupling strength $G_{AB}$ is
\be
   G_{AB}=\frac{2\wi+4}{\si(2\wi+3)}\left[1
   -\frac{1}{\wi+2}
   (s_{A}+s_{B}-2s_{A}s_{B})\right]
\ee
to 1PK order. (Note that the above equation is a
simplification that ignores effects due to the relative velocity of the
two bodies.) If body {\scriptsize $B$} is a
non-self-gravitating test particle then $s_{B}=0$ and body
{\scriptsize $A$} has a Kepler measured mass
\be
   M_{K}^{A}=\frac{FR_{AB}}{M_{I}^{B}}=\frac{(2\wi+4)M_{I}^{A}}
   {\si(2\wi+3)}\left(1-\frac{s_{A}}{\wi+2}\right).
\ee
In ST gravity, the $\Phi$ field outside an asymptotically flat static
black hole is constant and one can show that the sensitivity $s_{BH}$
of a black hole is $s_{BH}=\half$. Assuming that a small black hole in
orbit about body {\scriptsize $A$} also has this property, then the
orbital mass in this case is given by
\be
   \frac{(2\wi+4)M_{I}^{A}}{\si(2\wi+3)}
   \left(1-\frac{1}{2\wi+4}\right)=\frac{M_{I}^{A}}{\si}
\ee
which consistent with our relation (\ref{MIdef}) between the inertial
and tensor masses.

The formalism devised by Damour \& Esposito-Farese \cite{DEF1} for
tensor-multi-scalar theories is similar to the modified EIH formalism
except that it describes the motion of the system in terms of Einstein
frame variables. In place of $s$, the fundamental quantity used is a
coupling parameter $\alpha$ which, for a star of mass $M_{ADM}^{*}$ and
Einstein frame scalar charge $Q_{S}^{*}$, is defined by
\be\label{adef}
   \alpha:=\frac{Q_{S}^{*}}{M_{ADM}^{*}}
\ee
and measures the strength of the coupling between $\vp$ and the
non-scalar matter. In the weak field limit,
$\alpha\ra -1/\sqrt{2\wi+3}$. The equations of motion at the 1PK level
then depend upon the values of $\alpha$ and $d\alpha/d\vp$, evaluated
at $\vp=\vp_{\infty}$.

Damour \& Esposito-Farese (see Appendix A of \cite{DEF1}) have shown
that, by varying the Einstein frame Hamiltonian with respect to changes
in the asymptotic value of the scalar field $\vp$, one can derive the
following relation:
\be\label{alphadef}
   \frac{Q_{S}^{*}}{M_{ADM}^{*}}=\frac{1}{M_{ADM}^{*}}
   \frac{\partial M_{ADM}^{*}}{\partial\vp}
\ee
where the derivative is evaluated at $\vp=\vp_{\infty}$ for fixed
particle number.
A similar analysis, performed on the Jordan frame Hamiltonian, gives
\be\label{alts}
   \frac{Q_{S}}{M_{I}}=\frac{-2}{2\wi+3}(1-2s).
\ee
Using eqn (\ref{alts}) one can calculate $s$ directly from the
scalar charge and inertial mass. Combining eqns (\ref{alphadef}) and
(\ref{alts}) and using eqns (\ref{Mrelate}) and (\ref{MIdef}), we 
find that the parameters $\alpha$ and $s$ are related by
\be\label{as}
   \alpha=\frac{-1}{\sqrt{2\wi+3}}(1-2s),
\ee
to 1PK order.

Formally ST gravity reduces to GR in the limit $\omega\ra\infty$,
$\Phi\ra\Phi_{GR}=1$ (in units with $G=1$ and
provided $g^{ab}T_{ab}$ is non-zero). In this
limit, $s$ generally approaches a  non-zero value and measures the
response of $M_{I}$ to a (global) change in the value of Newton's
constant $G$.

The modified EIH formalism is adequate for describing the motion of
compact bodies when effects such as spontaneous scalarization are
absent. However, as shown in \cite{DamourEF,DEF2}, when scalarization
occurs in a neutron star, both $\alpha$ and $Q_{S}^{*}$ remain finite
and non-zero as $\wi\ra\infty$. Equation (\ref{as}) then implies that $s$
must diverge and so the modified EIH formalism breaks down. We shall
discuss this further in Section \ref{Equ}, where we also show how $s$
behaves for various boson star models.

\section{Boson Stars}

Boson stars are
gravitationally bound configurations of zero temperature bosons and
many of their properties in ST gravity have already been studied
\cite{Gund,Torres1,Comer,AW,AWDT}. They share many
features in common with simple neutron star and white dwarf solutions. 
Sequences of boson star solutions may be parameterised by
the central boson field density, and when plotted against this
parameter, the mass and conserved charge curves show a maximum at
which instability first occurs.

We take as a matter source a complex self-interacting boson field $\Psi$ with
Lagrangian
\be\label{lagrangian}
   L_{m}=-\half g^{ab}\left(\partial_{a}\Psi\partial_{b}{\overline\Psi}
   +\partial_{a}{\overline\Psi}\partial_{b}\Psi\right)
   -\Psi{\overline\Psi}-4\pi\Lambda({\overline\Psi}\Psi)^{2}
\ee
where $\Lambda$ measures the strength of the boson self-interaction 
and we have
chosen units in which $\hbar/\mu=1$, where $\mu$ is the boson mass.
The Jordan frame energy momentum tensor for this field is
\be
   T_{ab}=\partial_{a}\Psi\partial_{b}{\overline\Psi}
   +\partial_{a}\Psi\partial_{b}{\overline\Psi}-\half g_{ab}
   g^{cd}(\partial_{c}\Psi\partial_{d}{\overline\Psi}
   +\partial_{c}{\overline\Psi}\partial_{d}\Psi)
   -g_{ab}\Psi{\overline\Psi}-4\pi g_{ab}\Lambda({\overline\Psi}{\Psi})^{2}
\ee

We consider static, spherical solutions with the line element
(\ref{linel}) and 
we only consider minimum
energy solutions, which implies that the boson field wave function may
be written as
\be\label{psidef}
   \Psi=\frac{P(r)}{\sqrt{8\pi}}\exp(i\Omega t)
\ee
(see Appendix)
where $P(r)$ is a real dimensionless function and the constant $\Omega$
is real. For the metric given implicitly in the line element (\ref{linel}),
the independent components of the field and matter equations
(\ref{JFfeqns}) and (\ref{JFothereqns}) are
\begin{eqnarray}\label{dmdr}
   m^{\prime}=\frac{r}{2\omega+3}\left[(r-2m)
   \left(\frac{\Phi^{\prime 2}}{2\Phi}\frac{d\omega}{d\Phi}
   +\frac{P^{\prime 2}}{2\Phi}(1+2\omega)\right)
   +\frac{rP^{2}}{2\Phi}\left(2\omega-1+(2\omega+5)\frac{\Omega^{2}}{B}
   \right.\right.\nonumber \\ \left.\left.
   +(2\omega-1)\frac{P^{2}\Lambda}{2}\right)\right]
   +(r-2m)\left[\frac{r\Phi^{\prime}B^{\prime}}
   {4B\Phi}+\frac{\Phi^{\prime 2}r\omega}{4\Phi^{2}}\right],
\end{eqnarray}
\be\label{dbdr}
   B^{\prime}=\frac{1}{2\Phi+r\Phi^{\prime}}\left[\frac{\Phi^{\prime
   2}\omega B r}{\Phi}-4B\Phi^{\prime}+2P^{\prime 2}Br
   -\frac{1}{r-2m}\left(2P^{2}r^{2}(B-\Omega^{2})+P^{4}B\Lambda
   +\frac{4Bm\Phi}{r}\right)\right],
\ee
\be\label{ddsdrr}
   \Phi^{\prime\prime}=\Phi^{\prime}\left[\frac{rm^{\prime}-m}{r(r-2m)}
   -\frac{2}{r}-\frac{B^{\prime}}{2B}\right]-\frac{2}{2\omega+3}
   \left[\frac{\Phi^{\prime 2}}{2}\frac{d\omega}{d\Phi}+P^{\prime 2}
   +\frac{P^{2}r}{r-2m}\left(2-\frac{\Omega^{2}}{B}+2\Lambda P^{2}
   \right)\right]
\ee
and
\be\label{ddpdrr}
   P^{\prime\prime}=P^{\prime}\left[\frac{rm^{\prime}-m}{r(r-2m)}
   -\frac{2}{r}-\frac{B^{\prime}}{2B}\right]+\frac{Pr}{r-2m}
   \left[1-\frac{\Omega^{2}}{B}+\Lambda P^{2}\right],
\ee
where a prime denotes $d/dr$.
For the corresponding Einstein frame equations, see
\cite{Comer}. 

The existence of a global $U(1)$ symmetry in the matter Lagrangian
(\ref{lagrangian}) implies the existence of a conserved charge $N$. Using
the above coordinates and field variables, this may be written as
\be\label{Number}
   N=\integ \frac{r^{2}\Omega P^{2}}{\sqrt{B}\sqrt{1-2m/r}}\;dr
\ee
and is interpreted as the total number of bosons in the
star.

Equations (\ref{dmdr}) to (\ref{ddpdrr}) must be integrated
numerically. To ensure that the solutions describe bound eigenstates
of $\Psi$ one must impose the boundary condition
\be
   \lim_{r\to\infty}P(r)=0
\ee
and the regularity conditions
\be
   m_{0}=0,\;\;\;\;P^{\prime}_{0}=0,\;\;\;\;\Phi^{\prime}_{0}=0,
\ee
where the subscript `{\scriptsize $0$}' denotes values at the origin. The
field equations then become eigenvalue equations for $\Omega$, are
parameterised by $P_{0}$ for fixed $\si$ and automatically lead
to the boundary conditions $\Phi=\si+{\cal O}(r^{-1})$,
$m(r) = M_{ADM}+{\cal O}(r^{-1})$ and
$B = B_{\infty}+{\cal O}(r^{-1})$ as $r\ra\infty$. For any solution,
one can make the rescaling $B_{\infty}\ra 1$ by rescaling $\Omega$
appropriately.

In Brans-Dicke theory, where $d\omega/d\Phi=0$, the
equations with $\Lambda=0$ are invariant under the rescaling
\be\label{rescale1}
   P\ra {\kappa}P, \;\;\;\; \Phi\ra {\kappa}^{2}\Phi
\ee 
where ${\kappa}$ is a constant.
Equation (\ref{rescale1}) leaves $M_{T}$, $M_{K}$ and $M_{ADM}$
invariant and rescales the particle number as
\be\label{Nrescale}
   N\ra {\kappa}^{2}N.
\ee
We can use this scaling invariance to find an alternative expression
of $s$ in a $\Lambda=0$ BD boson star
as follows. Consider a pair of solutions,
$\sigma_{1}$ and $\sigma_{2}$, with the same boundary value $\si$ and
with central boson field amplitudes
$P_{0}^{(1)}$ and $P_{0}^{(2)}=P_{0}^{(1)}+\delta P_{0}$, such that
$\delta P_{0}\ll 1$. Solution $\sigma_{1}$ has mass $M_{T}^{(1)}$ and
particle number $N^{(1)}$, while $\sigma_{2}$ has mass $M_{T}^{(2)}$
and particle number $N^{(2)}$. Then, to first order in $\delta
P_{0}$,
\be\label{mas}
   M_{T}^{(2)}=M_{T}^{(1)}+\frac{\partial M_{T}}{\partial P_{0}}\delta P_{0}
\ee
and
\be\label{num}
   N^{(2)}=N^{(1)}+\frac{\partial N}{\partial P_{0}}\delta P_{0},
\ee
where the derivatives are taken with $\si$ held fixed. We
use eqn (\ref{rescale1}) to generate a new solution ${\wt\sigma}_{2}$
with mass ${\wt M}_{T}^{(2)}$, particle number ${\wt N}^{(2)}$ and
boundary scalar field ${\wt\Phi}_{\infty}$ and we choose a value of
${\kappa}$ such that ${\wt N}^{(2)}:={\kappa}^{2}N^{(2)}=N^{(1)}$. From eqn
(\ref{num}), ${\kappa}^{2}$ must then be given by
\be
   {\kappa}^{2}=1-\frac{1}{N^{(1)}}\frac{\partial N}{\partial P_{0}}\delta
   P_{0}
\ee
to first order in $\delta P_{0}$. This implies that
\be\label{newS}
   {\wt\Phi}_{\infty}={\kappa}^{2}\si=\si\left(1-\frac{1}{N^{(1)}}
   \frac{\partial N}{\partial P_{0}}\delta P_{0}\right),
\ee
while $M_{T}^{(2)}$ is invariant under
this rescaling, thus ${\wt M}_{T}^{(2)}=M_{T}^{(2)}$.
Then, using eqn (\ref{mas}), we have
\be
   \delta M_{T}:={\wt M}_{T}^{(2)}-M_{T}^{(1)}=
   \frac{\partial M_{T}}{\partial P_{0}}\delta P_{0}
\ee
and, from eqn (\ref{newS}),
\be\label{deltasi}
   \delta\si:={\wt\Phi}_{\infty}-\si=
   -\frac{\si}{N^{(1)}}\frac{\partial N}{\partial P_{0}}\delta P_{0}.
\ee
Combining these results with the definition (\ref{MIdef}) gives
\be
   \frac{\delta M_{I}}{\delta\si}=\frac{\delta(\si M_{T})}{\delta\si}
   =M_{T}^{(1)}-\si\frac{\partial M_{T}}{\partial P_{0}}\delta P_{0}
   \left[\frac{\si}{N^{(1)}}\frac{\partial N}{\partial P_{0}}
   \delta P_{0}\right]^{-1}
\ee
\be
   =M_{T}^{(1)}-N^{(1)}\frac{\partial M_{T}}{\partial N}.
\ee
Taking the limit $\delta\si\ra 0$ and using eqn (\ref{sdef}) we
have
\be\label{BDs1}
   s=\frac{\si}{M_{I}}\left(M_{T}-N\frac{\partial M_{T}}
   {\partial N}\right)=1-\frac{N}{M_{T}}\frac{\partial M_{T}}
   {\partial N},
\ee
where we have dropped the label from $N$ and $M_{T}$ since this
relation is true for any initial solution $\sigma_{1}$. 
One can show \cite{AW} that for fixed $\si$ 
\be
   \frac{\partial M_{T}}{\partial N}=\frac{\Omega}{\si}.
\ee
Combining this result with eqn (\ref{BDs1}) gives an alternative
expression for $s$:
\be\label{BDs2}
   s=1-\frac{\Omega N}{\si M_{T}},
\ee
Note that eqns (\ref{BDs1}) and (\ref{BDs2})
hold for any value of $\omega$.

When $\Lambda$ is non-zero but finite, one cannot drive a similar
result since one also needs to rescale $\Lambda$ to keep the field
equations invariant. However, Gunderson \& Jensen \cite{Gund}
have shown that in
BD theory, even a relatively small value of $\Lambda$ causes the
terms that are quartic in $P$ to dominate the energy momentum tensor.
For $\Lambda>100$ the solutions are well approximated by taking the
limit $\Lambda\ra\infty$, and one can derive an expression for the
sensitivity of a boson star in this limit.
One first rewrites the field equations by making 
the field and coordinate re-definitions $\Pi=\sqrt{\Lambda}P$,
$\rho=r/\sqrt{\Lambda}$ and then takes the limit
$\Lambda\ra\infty$. The resulting equations are identical to
eqns (\ref{dmdr}) to (\ref{ddpdrr}) except that $P$ and $r$ are
replaced by $\Pi$ and $\rho$, $\Lambda=1$, all terms including a
factor of $\Pi^{\prime}$ vanish and $\Pi$ is given by the algebraic
expression 
\be
   \Pi^{2}=\frac{\Omega^{2}}{B}-1.
\ee 
See \cite{Gund} for details in BD theory.
The equations defining $N$, $Q_{S}$ and
$M_{T}$ are identical to the finite $\Lambda$ equations, except that
$P$ and $r$ are everywhere replaced by $\Pi$ and $\rho$. Then masses 
are measured in
units of $M_{pl}^{3}/\mu^{2}$ and in BD theory the new field
equations are invariant under the rescaling
\be\label{rescale2}
   \Phi\ra {\kappa}^{2}\Phi,\;\;\;\;\rho\ra {\kappa}\rho,
\ee 
where $\kappa$ is constant.
Under eqn (\ref{rescale2}), the mass and
particle number rescale as
\be\label{MNrescale}
   M_{T}\ra {\kappa} M_{T},\;\;\;\;N\ra {\kappa}^{3}N.
\ee
Performing a similar analysis to the one given above for the
$\Lambda=0$ case and using eqns (\ref{rescale2}) and (\ref{MNrescale}),
one can show that the sensitivity in the limit
$\Lambda\ra\infty$ is given by
\be\label{BDsLinf}
   s=\frac{3}{2}\left(1-\frac{N}{M_{T}}\frac{\partial M_{T}}{\partial N}
   \right)=\frac{3}{2}\left(1-\frac{N\Omega}{\si M_{T}}\right).
\ee
The factor of $3/2$ appears in these equations for two reasons:
firstly, in the
present case, $delta\Phi$ differs from eqn (\ref{deltasi}) by a factor
of $2/3$ (since the particle number $N^{*}$ scales differently from
$N^{*}$ by a factor of $\kappa^{3/2}$) and, secondly, one must
include an extra term in $\delta M_{T}^{*}$ to account for the
rescaling of $M_{T}^{*}$ in eqn (\ref{MNrescale}), a term which does
not appear for the $\Lambda=0$ case.

The first of these equalities is identical to the result derived for
neutron stars \cite{Eardley} (in the latter case, 
$N$ must be replaced by the neutron star's
conserved baryon number). This is because both neutron
stars and boson stars in the limit $\Lambda\ra\infty$ have masses that scale
as $M_{pl}^{3}/\mu^{2}$, where for the neutron star $\mu$ is the
baryon mass. in contrast, the mass of a boson star with no
self-interaction scales as $M_{pl}^{2}/\mu$, so there is no factor of
$3/2$ in eqns (\ref{BDs1}) and (\ref{BDs2}).

One cannot carry out the same analysis for more general ST theories
since the ST versions of eqns (\ref{rescale1}) and
(\ref{rescale2}) require $\omega(\Phi)$ to be held
fixed. Since $\Phi$ rescales under eqn (\ref{rescale1}), 
the functional form of $\omega$ must
change to compensate in a way that depends upon the precise choice of
$\omega$. However, eqns (\ref{BDs1}) and (\ref{BDsLinf}) will still be
approximately correct for solutions in which $d\omega/d\Phi$ remains small.

\section{Equilibrium Solutions}\label{Equ}

We have numerically integrated the field equations 
(\ref{dmdr}) to (\ref{ddpdrr}) for BD boson stars and for ST boson
stars with the coupling functions given by eqns (\ref{JFexp}) and
(\ref{nlcoupling}). A feature shared by all
solutions sets it that both $M_{T}$  and $N$ increase from zero at
$P_{0}=0$ to reach coinciding maxima $M_{T}^{(max)}$ and $N^{(max)}$
at some value of $P_{0}=P_{0}^{(max)}$. For $P_{0}>P_{}^{(max)}$,
the solutions become unstable. The magnitude of the
scalar charge $|Q_{S}|$
also increases from zero at $P_{0}=0$ but its first maximum
$|Q_{S}|^{(max)}$ does not
occur at the same value of $P_{0}$ as the maxima in $N$ and $M_{T}$.
For all solutions discussed here, we have found that $s<0.5$ which
implies that $Q_{S}$, $\alpha$ and $Q_{S}^{*}$ are all negative.

We consider first boson stars in BD theory. One can show that in the
weak field limit, where the (dimensionless) parameter
$P_{0}\ra 0$, the sensitivity is given by
\be\label{wfsBD}
   s=-2{\cal E}+{\cal O}(P_{0}^{2}),
\ee
where ${\cal E}$ is the fractional binding energy defined by
as
\be\label{Benergy}
   {\cal E}:=\frac{\si M_{T}-N}{N}.
\ee
For stable stars, ${\cal E}<0$ and hence all stable BD stars have
$s>0$ in the weak field limit. Numerical calculations then show that 
$s$ increases with $P_{0}$ to reach
some maximum at a value of $P_{0}>P_{0}^{(max)}$. Values of $s_{(max)}$,
the sensitivity of the maximum mass stable solution at
$P_{0}=P_{0}^{(max)}$, along with corresponding values
$\alpha_{(max)}$ of the coupling parameter $\alpha$, are shown 
in Table \ref{table1} for several
choices of $\omega$ for  both $\Lambda=0$ and in the limit $\Lambda\ra\infty$.

\begin{table}[ht]
\centering
\begin{tabular}{|c|c|c|c|c|c|}\hline
\multicolumn{3}{|c|} {$\Lambda\ra\infty$} & 
\multicolumn{3}{c|}{$\Lambda=0$}
\\ \hline
$\omega$ & $s_{(max)}$ & $\alpha_{(max)}$ & 
$\omega$ & $s_{(max)}$ & $\alpha_{(max)}$\\ \hline
1    &   0.211 & -0.258 & 1    &  0.144 & -0.318 \\
10   &   0.189 & -0.130 & 10   &  0.126 & -0.156 \\
500  &   0.182 & -0.020 & 500  &  0.121 & -0.024 \\
3300 &   0.178 & -0.008 & 3300 &  0.119 & -0.009 \\ 
\hline
\end{tabular}
\caption{\label{table1}Sensitivity $s$ and coupling parameter $\alpha$
of maximum mass boson stars in BD
gravity for $\Lambda=0$ and in the limit $\Lambda\ra\infty$.}
\end{table}     

As eqns (\ref{BDs2}) and (\ref{BDsLinf}) suggest, for each $\omega$
the sensitivity of a maximal mass
boson star with $\Lambda\ra\infty$ is greater than the
sensitivity of the corresponding
$\Lambda=0$ star by a factor of 3/2. This is also true for values of
$N$ less that the maximum value. For both choices
of $\Lambda$, $s_{(max)}$ decreases with increasing $\omega$ towards
some non-zero limit as $\omega\ra\infty$ and to the level of
accuracy quoted in the Table, $s_{(max)}$ for a boson star in GR has the
same value as for a $\omega=3300$ star. For comparison, maximum 
mass neutron stars
have been found to have sensitivities that vary from $s_{(max)}=0.2$ to
$s_{(max)}=0.39$, depending upon the equation of state chosen
\cite{Eardley,DEF2}. Hence for boson stars with no self-interaction,
the response of $M_{I}$ to changes in $\Phi$ is much smaller than for
neutron stars, while for boson stars with $\Lambda\ra\infty$, the
sensitivity is almost comparable to that of a neutron star. Note that, 
from the relation (\ref{as}), we have $\alpha\ra 0$ 
as $\omega\ra\infty$ since $s$ remains finite in this limit for all BD
solutions.

We next consider boson stars in more general ST theories. One can show 
that, for any ST theory and for any value of $\Lambda$, in 
the weak field limit $s$ is given by
\be\label{wfsST}
   s=-{\cal E}(2+(2\wi+3){\wt\beta})+{\cal O}(P_{0}^{2}).
\ee
From eqns (\ref{PPNpars}) and the limits (\ref{limits}), we have
for any weak field solution $s<0$ when ${\wt\beta}<-0.0003$.

Figure \ref{fig1} shows sensitivity $s$ against particle number $N$ for
several sets of boson star solutions with $\Lambda=0$, for the
coupling function (\ref{JFexp}) with the parameter choices $k=1$
and $k=3$. For this coupling function, ${\wt\beta}=-2k$ and the
inequality (\ref{betalimit}) is satisfied when $k>2$. For the values 
of $k$ we are using, it turns out that the first inequality of eqns
(\ref{explimits}) places the tightest constraints on the value of
$\si$ and these are
$(\si-1)\leq 7.5\ex^{-5}$ for $k=1$ and $(\si-1)\leq 8.3\ex^{-6}$ for
$k=3$. This latter value corresponds to the limit
$\vp_{\infty}\leq 0.0012$ which is smaller that the limit
$\vp_{\infty}\leq 0.0043$ quoted in \cite{DamourEF} since the
constraint on the PPN parameter $\gamma$ has recently been
tightened \cite{Will2}. For the sake of comparison, in the Figure we
have included $k=3$ solutions with the same boundary value used in
\cite{DamourEF}, although we also choose values compatible with the
current observational constraints.

The curves in Figure \ref{fig2} show the coupling parameter $\alpha$
against $N$ for the $k=3$ solutions shown in Figure \ref{fig1}. In
contrast with the corresponding neutron stars curves shown in
\cite{DamourEF}, the maximum value of $-\alpha$ occurs when
$P_{0}>P_{0}^{(max)}$ or, equivalently, when $N<N^{(max)}$ and the
solutions are unstable. This is because, for any set of boson star solutions,
$|Q_{S}|^{(max)}$ occurs at a parameter value $P_{0}>P_{0}^{(max)}$.

\begin{table}[ht]
\centering
\begin{tabular}{|c|c|c|c|c|c|c|c|c|}\hline
$k$ & ${\wt \beta}$ & $\Phi_{\infty}-1$ & $\vp_{\infty}$ & 
$\omega_{\infty}$ & $s_{(max)}$ & $\alpha_{(max)}$ & $-Q_{S}^{(max)}$
& $-Q_{S}^{*)(max)}$\\ 
\hline
1 & -2 & $7.5\ex^{-5}$ & $4.3\ex^{-3}$ & 6759 &
-0.0035 & -0.00920 & $1.0\ex^{-4}$ & 0.0058\\
3 & -6 & $1.1\ex^{-4}$ & $4.3\ex^{-3}$ & 750  & -2.70 &
-0.165 & $5.3\ex^{-3}$ & 0.10\\
3 & -6 & $1.1\ex^{-6}$ & $4.3\ex^{-4}$ & 75114 &
  -4.00 &  -0.0232 & $7.6\ex^{-5}$ & 0.015\\
3 & -6 & $1.1\ex^{-8}$ & $4.3\ex^{-5}$ &  
$7.51\ex^{6}$ & -4.02   & -0.00233 & $7.6\ex^{-7}$ & 0.0015 \\
3 & -6 & $1.1\ex^{-10}$ & $4.3\ex^{-6}$ & 
$7.51\ex^{8}$ & -4.02   & -0.000233 & $7.6\ex^{-9}$ & 0.00015 \\
\hline
\end{tabular}
\caption{\label{table2}Numerical data for boson stars with $\Lambda=0$
and $2\omega+3=1/(2k\log\Phi)$. The label {\scriptsize $(max)$}
denotes values at the maximum mass solution.}
\end{table}     

Some of the data taken from the numerical calculations are shown in Table
\ref{table2}. Together with Figures \ref{fig1} and \ref{fig2}, they
show that the scalarization phenomenon does not occur in $\Lambda=0$
boson stars. As the study in
\cite{DamourEF} shows, in the limit $(\si-1)\ra 0$ and for neutron stars 
with ${\wt\beta}<-4$, the values of both $-\alpha$  and 
$Q_{S}^{*}$ for a star of given baryon number approach
some finite limit (which is either zero for stars whose mass is below
some critical value or non-zero for stars whose mass is above this value).
From eqn (\ref{as}), this implies that, for stars above the critical mass,
$s$ is negative and also that $s$ diverges as $\sqrt{\wi}$ in this
limit. In addition, from eqn (\ref{QSrelate}), $Q_{S}$ vanishes as
$1/\sqrt{\wi}$. In contrast, for boson stars without self-interaction
we find that for all solutions $s$ reaches some finite, non-zero limit as
$(\si-1)\ra 0$. This implies that $\alpha$ must vanish in this limit
for all solutions, as shown by eqn (\ref{as}). As shown in 
Table \ref{table2}, the magnitude of both $Q_{S}^{(max)}$ and 
$Q_{S}^{*(max)}$ decrease to zero as $\si-1\ra 0$ and this is true in
these solutions for all other values of $N$.

We next consider $\Lambda=0$ boson stars with the coupling function
(\ref{nlcoupling}). Assuming that $\si -1$ is small, we have
${\wt\beta}=-4\xi$ which implies that the inequality (\ref{betalimit})
is satisfied when $\xi>1$. 
Figures \ref{fig3} and \ref{fig4} show curves of $s$ and $\alpha$
against $N$ for solutions with $\xi=1$ and $\xi=2$. For these choices
of $\xi$, the first of eqns (\ref{nllimits}) places the tightest
limits on the values of $\si$, which are
$(\si-1)<3.75\ex^{-5}$ for $\xi=1$ and $(\si-1)<1.87\ex^{-5}$ for
$\xi=2$. In the Figures we have chosen boundary
values compatible with these limits. The behaviour of the solutions
is qualitatively similar to the behaviour of the solutions shown in
Figures \ref{fig1} and \ref{fig2} and, again, the maximum value of
$-\alpha$ occurs after $N$ has reached its first maximum. Data taken
from these calculations is shown in Table \ref{table3}.

\begin{table}[ht]
\centering
\begin{tabular}{|c|c|c|c|c|c|c|c|}\hline
$\xi$ & ${\wt \beta}$ & $\Phi_{\infty}-1$ & $\omega_{\infty}$ & $s_{(max)}$ &
$\alpha_{(max)}$ & $-Q_{S}^{(max)}$ & $-Q_{S}^{*(max)}$\\ \hline
1 &-4   & $3.75\ex^{-5}$  & 3329  &  -0.433  & -0.0229 & 0.00035 & 0.014\\
1 &-4   & $1.87\ex^{-5}$  & 6679  &  -0.438  & -0.0162 & 0.00018 & 0.010\\
1 &-4   & $1.91\ex^{-6}$  & 65536 &  -0.422  & -0.0051 &0.000018 & 0.0032\\
2 &-8   & $1.87\ex^{-5}$  & 3339  &  -7.39   & -0.193  & 0.0028  & 0.113\\
2 &-8   & $9.41\ex^{-6}$  & 6637  &  -10.1   & -0.184  & 0.0019  & 0.109\\
2 &-8   & $2.38\ex^{-7}$  &262144 &   -45.7  & -0.128  & 0.00021 & 0.077\\
\hline
\end{tabular}
\caption{\label{table3}Numerical data for boson stars with $\Lambda=0$
and $\omega=\Phi/(8\xi(\Phi-1))$.}
\end{table}     

Finally, we consider solutions to the field equations in the limit
$\Lambda\ra\infty$ for the coupling function (\ref{JFexp}) and for
$k=3$. As mentioned above, solutions with this limiting value of
$\Lambda$ are a good approximation of solutions having large but
finite values of $\Lambda$ in BD theory, and we expect this to be true
for other ST theories. Figure \ref{fig5} shows curves of $\alpha$
against $N$ for these solutions and data
taken from these integrations is shown in Table \ref{table4}.
In contrast with the $\Lambda=0$ solutions,
these solutions show the same scalarization phenomena as the neutron
stars studied in \cite{DamourEF,SalSud}. When the boundary value
$(\si-1)>0$, the coupling parameter $\alpha$ is small and negative
for small $N$ and decreases smoothly as $N$ increases to its first
maximum. For these solutions, $\wi$ is finite and all have finite, but
negative, sensitivities. In the limit $(\si-1)\ra 0$, $\wi\ra\infty$,
the solutions divide into two classes: for those with $N$ below a
critical value $N_{c}=0.200$, $\alpha$ vanishes
while $s$ decreases smoothly with increasing $N$
to diverge at $N=N_{c}$. For stars with
$N>N_{c}$, $\alpha$ is non-zero and increases rapidly with $N$, while
$s$ remains divergent. The star with particle number $N_{c}$ mass
$M_{T}=0.189$. Note that Figure \ref{fig5} shows that the
transition point between small and large values of $-\alpha$ becomes
sharper as $(\si-1)$ decreases, just as is the case for neutron stars.

\begin{table}[ht]
\centering
\begin{tabular}{|c|c|c|c|c|c|c|c|}\hline
$\Phi_{\infty}-1$ & $\omega_{\infty}$ & $s_{(max)}$ &
$\alpha_{(max)}$ & $-Q_{S}^{(max)}$ & $-Q_{S}^{*(max)}$\\ \hline
$1.1\ex^{-4}$  & 750            & -6.96 & -0.385 & $3.51\ex^{-3}$ & 0.068\\
$1.1\ex^{-8}$  & $7.51\ex^{6}$  &  -141  & -0.073 & $7.09\ex^{-6}$ & 0.014\\
$1.1\ex^{-10}$  & $7.51\ex^{8}$  &  -1125  & -0.058 & $5.66\ex^{-7}$ &0.011\\
$1.1\ex^{-12}$  & $7.51\ex^{10}$ &  -10829  & -0.056  & $5.45\ex^{-8}$&0.011\\
\hline
\end{tabular}
\caption{\label{table4} Numerical data for boson stars in the limit
$\Lambda\ra\infty$ with $2\omega+3=1/(2k\log\Phi)$ and for $k=3$.}
\end{table}

\section{Discussion}

We have analysed spontaneous scalarization is ST bosons stars for
several choices of the coupling function $\omega$. We have found that
scalarization does not occur when the bosons have no
self-interaction since, in this case, the sensitivity (or compactness) of the
stars is small. With the inclusion of a large quartic
self-interaction, the stars become much more compact and spontaneous
scalarization occurs. The fact that this
phenomenon occurs for boson stars as well as for simple neutron
star models suggests that scalarization may be a universal
characteristic of ST gravity. 

We have also given a brief introduction to the modified EIH formalism,
in which the sensitivity plays an important role, and shown how one
can calculate the sensitivity of a boson star (or any other compact
object) in ST theory from its mass and scalar charge. As well as
giving an indication of the compactness of a star, the sensitivity
also measures the response of the star's inertial mass to changes in
the asymptotic value of $\Phi$. Hence some of our results may also be
applied to the study of gravitational evolution, in which the
structure, and in particular the mass, of a star embedded in a
cosmological background is forced to evolve as the value of the
cosmological scalar field changes with cosmological time.
This phenomenon was first studied for ST
boson stars by Comer \& Shinkai \cite{Comer} and for BD boson stars by
Torres, Schunck \& Liddle \cite{TSL}. These authors assumed that
the star evolved quasi-statically and could be modelled as a sequence
of asymptotically flat equilibrium solutions of constant particle
number $N$ whose boundary value
$\si$ matches the cosmological value of $\Phi$ (which in general is an
increasing function of cosmological time). These results were extended
in \cite{AWDT}. Gravitational evolution has
also been shown to affect the cooling rate of white dwarfs, which
provides a new method of constraining ST theories of gravity \cite{BAT}.

Given a value of $s$ for a star one immediately knows how its
inertial mass will evolve, given the assumptions made in
\cite{Comer,TSL,AWDT}. For example, in BD theory we have found that $s$ is
positive for all boson star solutions, and is positive for all BD
white dwarf and neutron star solutions that we are aware of. Hence
$M_{I}$ will be an increasing function of cosmological time for all
of these objects. Since $M_{I}=\si M_{T}$, the fractional binding
energy equation ({\ref{Benergy}) shows that a BD boson star that is
stable at the current epoch will have been stable at all earlier
epochs, when $M_{I}$ would have had a smaller value.

For other ST theories $s$ may be negative so that
$M_{I}$ will be a decreasing function of time. This was indeed the
case for the boson stars studied by Comer \& Shinkai in \cite{Comer},
who modelled the evolution in the Einstein frame using a coupling
function equivalent to eqn (\ref{EFexp}). As we have seen, $s<0$ for
this coupling and the mass $M_{I}$ of the star will be larger at
earlier epochs. As pointed out in \cite{Comer},
this implies that the fractional binding energy
increases as we move further into the past and will eventually become
positive, at which point the star becomes unstable. Thus, knowing the
sign of $s$ for a boson star, one can immediately see if can exist as
a stable object at earlier cosmological epochs. These comments
apply equally well to other stellar objects.

The tensor mass (as we have defined it here) will
in general decrease with time for all ST theories. From the
definitions (\ref{sdef}) and (\ref{MIdef}) we have
\be\label{MTevolve}
   \frac{\partial M_{T}}{\partial\Phi}=\frac{M_{I}}{\si}(s-1)
\ee
where the derivative is evaluated at $\Phi=\si$. For a black hole
$s=0.5$ and, since a black hole is the most compact of all stellar
objects, one would expect other bodies to have $s<0.5$. This is
certainly the case for the boson star solutions discussed here. Then
eqn (\ref{MTevolve}) implies that $M_{T}$ will always be a decreasing
function of time in any cosmological model in which
$\Phi$ increases with cosmological time.

\appendix
\section{Appendix: Minimum Energy Solutions}

A proof that eqn (\ref{psidef}) leads to minimum energy solutions for
static, spherically symmetric boson stars can
be found in \cite{FLP1}.
However, as noted by Jetzer \cite{reviews1}, no corresponding
proof has been given for the case of a boson field with $U(1)$ charge. For
completeness, we generalise the proof in \cite{FLP1} to include charge
and to allow for arbitrary static, asymptotically flat spacetimes.
   
We start with the matter Lagrangian $L_{m}$ for bosons carrying charge
$e$, which reads
\be\label{Lag}
   L_{m}=-\half g^{ab}\left({\overline D}_{a}{\overline\Psi}D_{b}\Psi
   +D_{a}\Psi {\overline D}_{b}{\overline\Psi}\right)-{\overline\Psi}\Psi
   -V({\overline\Psi}\Psi)-\frac{1}{4}F_{ab}F^{ab},
\ee
where 
\be
   F_{ab}=\partial_{b}{\cal A}_{a}-\partial_{a}{\cal A}_{b}
\ee
is the Faraday tensor with ${\cal A}_{a}$ the vector potential,
\be
   D_{a}:=\partial_{a}+ie {\cal A}_{a}
\ee
is a covariant derivative operator, $V$ is the boson
self-interaction term (which we assume has no explicit time dependence
but is otherwise left arbitrary) and the over bar
denotes complex conjugation. We decompose the
spacetime into spacelike slices $\Sigma$ orthogonal to a timelike
Killing vector field $\xi^{a}$ and write the line element as
\be\label{slinel}
   ds^{2}=-B(x^{i})dt^{2}+h_{ij}(x^{i})dx^{i}dx^{j}
\ee
where $i,j$ label spatial indices, $x^{i}$ are coordinates on $\Sigma$
and $\xi^{a}$ has components $\xi^{a}=(1,0,0,0)$.
One can choose a gauge in which 
\be
   {\cal A}_{a}=({\cal A}_{0}(x^{i}),0,0,0)
\ee
and one can show that, for the Lagrangian (\ref{Lag}), the
conserved particle number may be written as
\be\label{Ndef}
   N=\int_{\Sigma}\sqrt{\frac{h}{B}}\left[i(\Psi\partial_{t}{\overline\Psi}
   -{\overline\Psi}\partial_{t}\Psi)+2e{\cal A}_{0}
   {\overline\Psi}\Psi\right]d^{3}x
\ee
where the subscript `{\scriptsize 0}' denotes the time component and 
$h=$Det$(h_{ij})$.

The boson field $\Psi$ has generalised momentum and velocity
\be
   p=\sqrt{-g}\frac{\partial L_{m}}{\partial(\partial_{t}\Psi)},\;\;\;\;
   \partial_{t}q=\partial_{t}\Psi
\ee
while the momentum and velocity associated with ${\overline\Psi}$ are
${\overline p}$ and ${\overline{\partial_{t}q}}$. Since we are
assuming that the
spacetime is static, the
Hamiltonian $H$ for the system is given by
\be
   H=\int_{\Sigma}d^{3}x\left(p\partial_{t}q+{\overline p}
   {\overline{\partial_{t}q}}\right) 
   -\frac{\partial I}{\partial t}+H_{s}
\ee
where $I$ is given by eqn (\ref{JFaction}) using the matter Lagrangian
(\ref{Lag}) and $H_{s}$ is a surface term that is independent of
$\Psi$ and ${\overline\Psi}$. In an asymptotically flat spacetime,
$M_{T}=H\si^{-1}$. 

We consider a sequence of solutions to the field equations (\ref{JFfeqns}),
with $T_{ab}$ derived from the Lagrangian (\ref{Lag}), in which the fields
$g_{ab}$, $\Phi$, $\Psi$ and ${\overline\Psi}$ are held fixed while
$\partial_{t}{\overline\Psi}$ and $\partial_{t}\Psi$ are allowed to
vary. Equivalently, we vary $p$ and ${\overline p}$ while keeping the
conjugate variables $q$ and ${\overline q}$ and the gravitational
fields fixed.
The desired form of these time derivatives is then the one that
minimises the Hamiltonian, subject to the constraint that $N$ is conserved.
Formally, we have
\be\label{var1}
   \delta(H-\Omega N)=0
\ee
under variations $\delta(\partial_{t}\Psi)$ and 
$\delta(\partial_{t}{\overline\Psi})$, where $\Omega$ is a Lagrange 
multiplier. Since all other fields are being held fixed, we only need to
consider terms in $H$ and $N$ that involve time derivatives of the
boson fields. These terms are
\be
   H_{\Psi}=\int_{\Sigma}d^{3}x\sqrt{\frac{h}{B}}\left[
   \partial_{t}{\overline\Psi}(\partial_{t}\Psi+ie{\cal A}_{0}\Psi)
   +\partial_{t}\Psi(\partial_{t}{\overline\Psi}
   -ie{\cal A}_{0}{\overline\Psi})  
   -\partial_{t}{\overline\Psi}\partial_{t}\Psi
   -ie{\cal A}_{0}(\Psi\partial_{t}{\overline\Psi}
   -\Psi\partial_{t}{\overline\Psi})\right]
    \nonumber
\ee
\be
   =\int_{\Sigma}d^{3}x\sqrt{\frac{h}{B}}\partial_{t}\Psi
   \partial_{t}{\overline\Psi}
\ee
and
\be\label{Npsi}
   N_{\Psi}=\int_{\Sigma}d^{3}x\sqrt{\frac{h}{B}}i(\Psi\partial_{t}
   {\overline\Psi}-{\overline\Psi}\partial_{t}\Psi),
\ee
while eqn (\ref{var1}) is equivalent to 
\be\label{var2}
   \delta(H_{\Psi}-\Omega N_{\Psi})=0.
\ee
Substituting in the explicit forms of $H_{\Psi}$ and $N_{\Psi}$, eqn
(\ref{var2}) becomes
\be
   \int_{\Sigma}d^{3}x\sqrt{\frac{h}{B}}\left[
   \delta(\partial_{t}\Psi)\left(\partial_{t}{\overline\Psi}
   +i\Omega{\overline{\Psi}}\right)+\delta(\partial_{t}{\overline\Psi})
   \left(\partial_{t}\Psi-i\Omega\Psi\right)\right]=0
\ee
which can only be satisfied if the equations
\be
   \partial_{t}\Psi-i\Omega\Psi=0,\;\;\;\;
   \partial_{t}{\overline\Psi}+i\Omega{\overline\Psi}=0
\ee
hold. This implies that
\be
   \Psi=f(x^{i})\exp(\Omega t)
\ee
where $f$ is a real function.
This result holds for arbitrary $V$ (provided $V$ has no explicit time
dependence) and is independent of the form of the gravitational sector
of the Lagrangian. We note here that one can also derive
this result by including the variations $\delta\Psi$,
$\delta(\partial_{i}\Psi)$ and their complex conjugates by assuming
the wave equations for the boson fields are satisfied.


\newpage

\begin{figure}
\begin{center}
\includegraphics{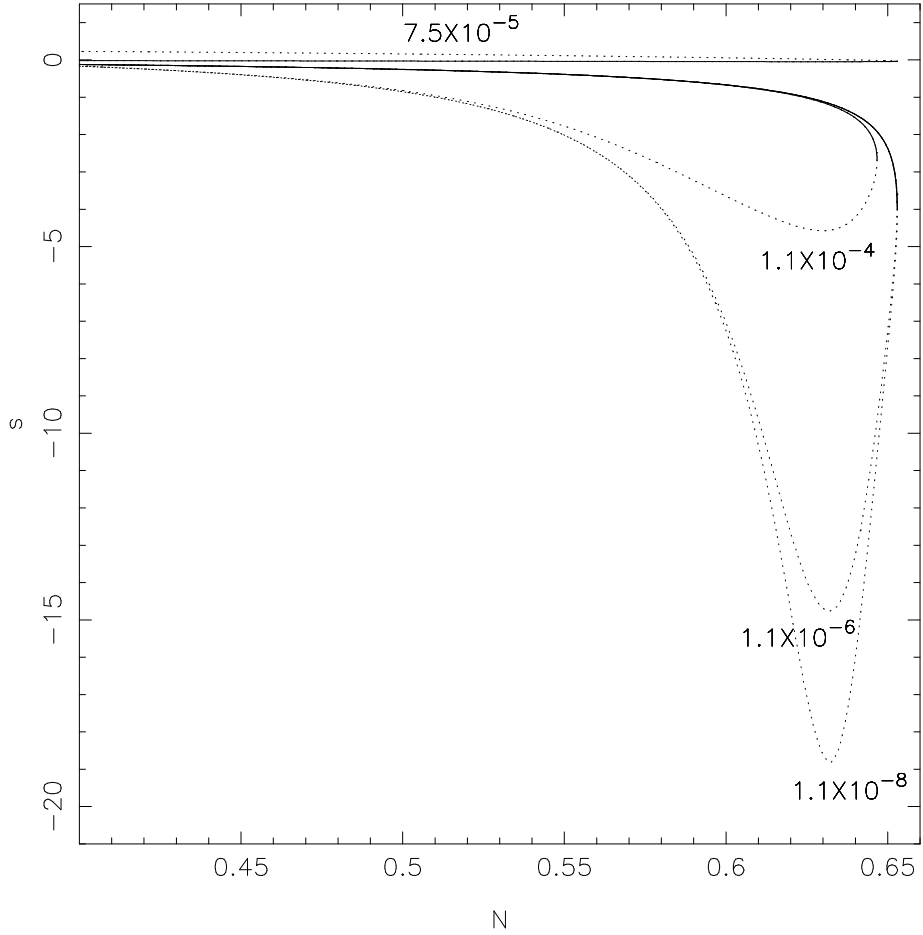}
\caption{\label{fig1}Sensitivity $s$ against particle number $N$ for ST
boson stars with $2\omega+3=1/(2k\log\Phi)$ and $\Lambda=0$. 
The curves are labelled by the value of
$(\si-1)$ and the solid portion of each represents stable
solutions. The uppermost curve corresponds to the parameter choice
$k=1$. The remaining three curves show solutions with $k=3$ and, at
the level of detail resolvable from the Figure, solutions with $k=3$
and $(\si-1)<1.1\ex^{-8}$ are indistinguishable from the lower most
curve shown.}
\end{center} 
\end{figure} 

\begin{figure}
\begin{center}
\includegraphics{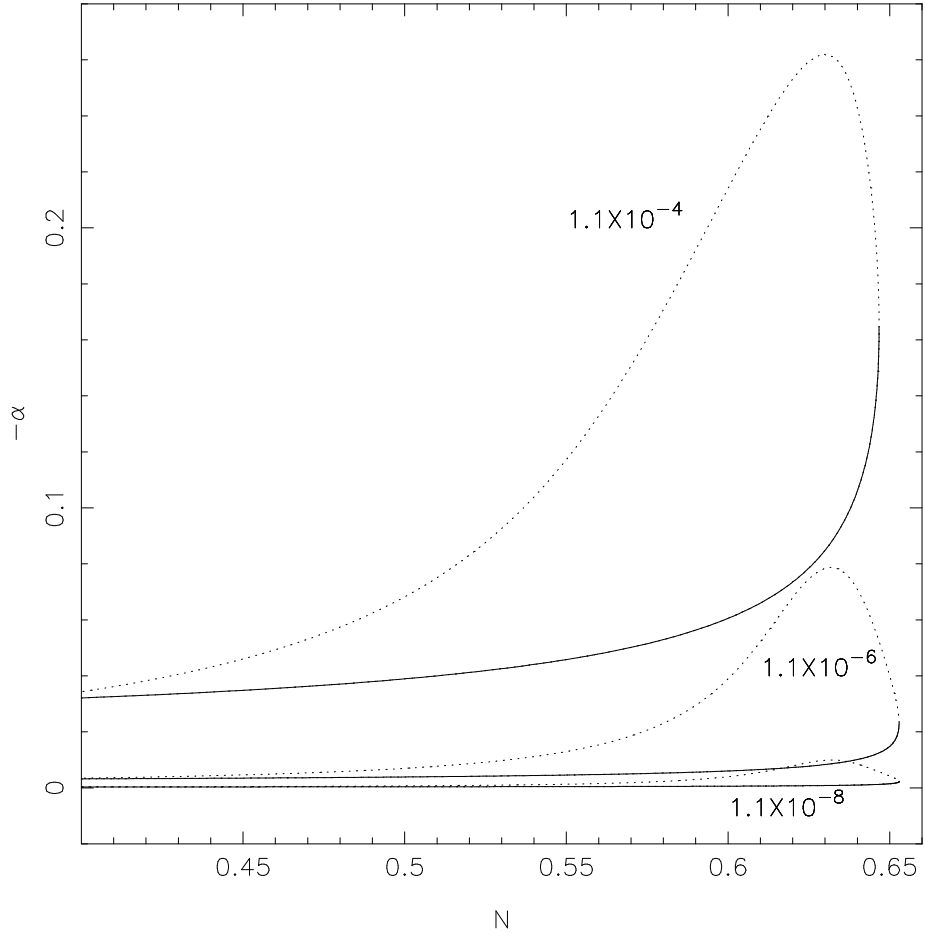}
\caption{\label{fig2}Coupling parameter $\alpha$ against particle 
number $N$ for ST boson stars with $2\omega+3=1/(2k\log\Phi)$ and
$\Lambda=0$. The curves are labelled by the value 
of $(\si-1)$ and all are for $k=3$. The solid
portion of each curve represents stable solutions. As $(\si-1)\ra 0$,
$-\alpha\ra 0$ for all solutions.}
\end{center} 
\end{figure} 

\begin{figure}
\begin{center}
\includegraphics{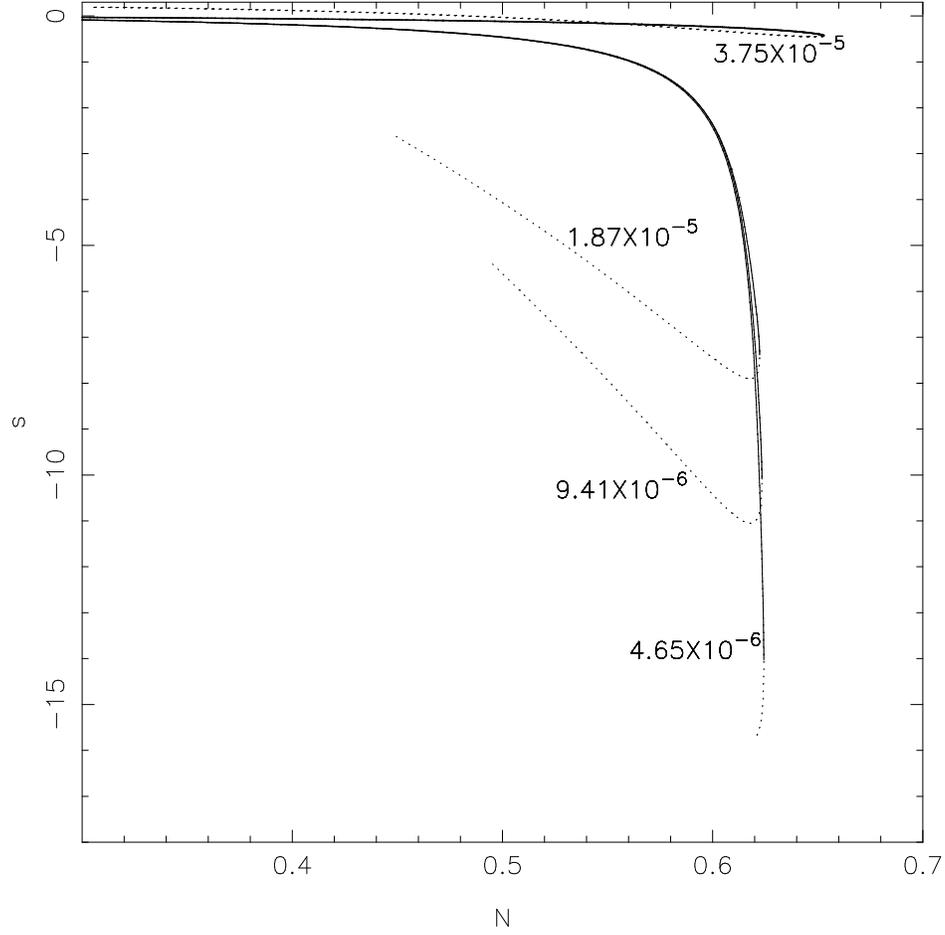}
\caption{\label{fig3}Sensitivity $s$ against particle number $N$ for ST
boson stars with $\omega=\Phi/(8\xi(\Phi-1))$ and $\Lambda=0$.
Curves are labelled by the value of $(\si-1)$ and the solid portion of
each represents stable solutions. The uppermost curve corresponds to
the parameter choice $\xi=1$ and is indistinguishable from $\xi=1$
curves with $(\si-1)<3.75\ex^{-5}$ at the level of resolution in the
Figure. The remaining three curves are for the parameter choice $\xi=2$.}
\end{center} 
\end{figure} 

\begin{figure}
\begin{center}
\includegraphics{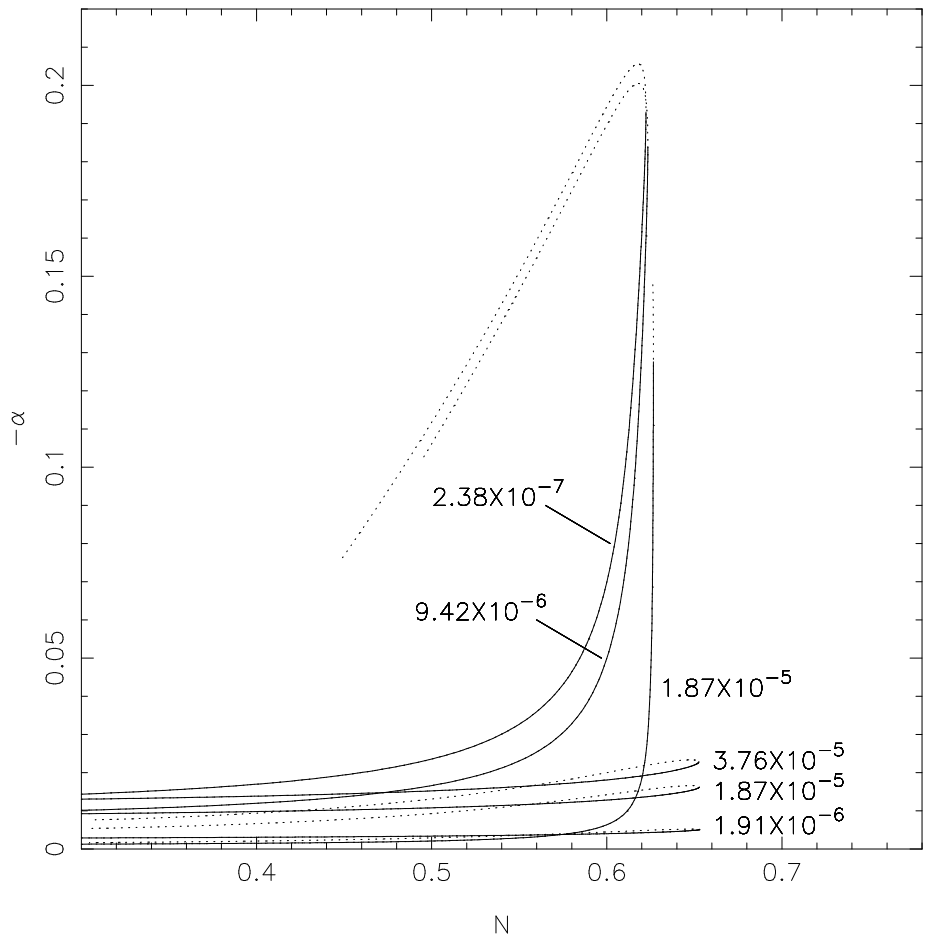}
\caption{\label{fig4}Coupling parameter $\alpha$ against particle 
number $N$ for ST boson stars with $\omega=\Phi/(8\xi(\Phi-1))$ and
$\Lambda=0$. The curves are labelled by the value of $(\si-1)$ and
the solid portion of each represent stable solutions. The lower three
curves are for the parameter choice $\xi=1$ while the remaining three
are for $\xi=2$.}
\end{center} 
\end{figure} 

\begin{figure}
\begin{center}
\includegraphics{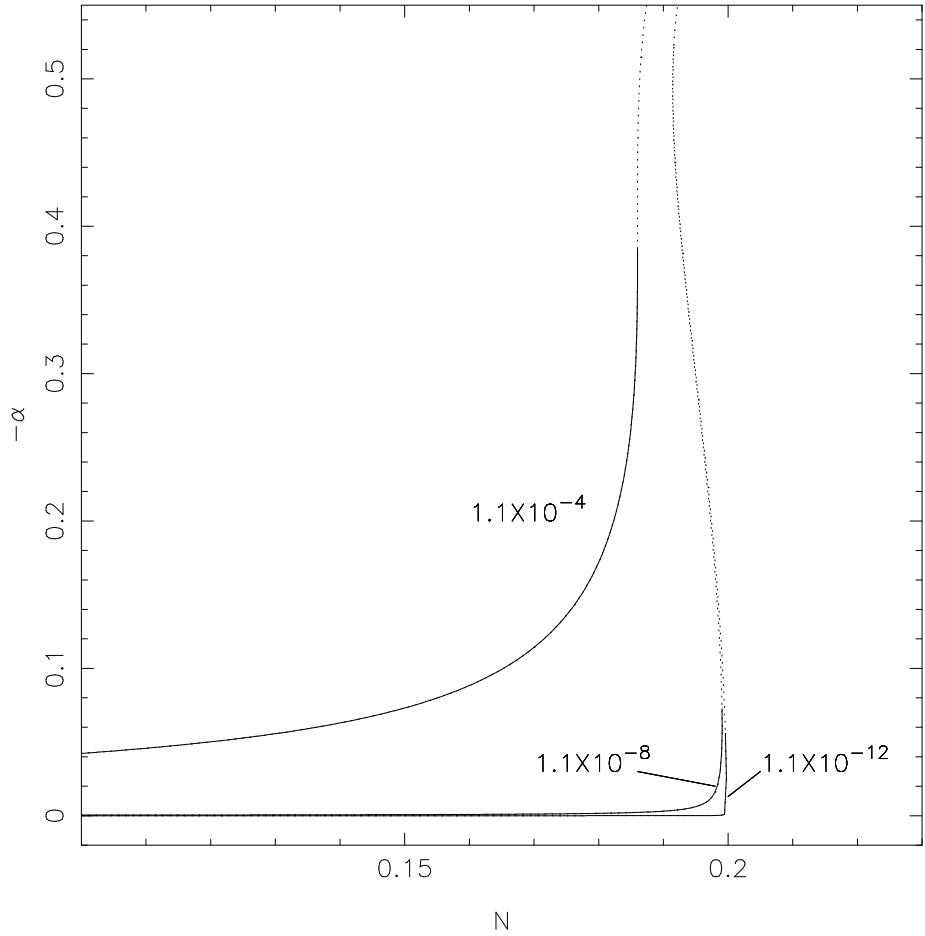}
\caption{\label{fig5}Coupling parameter $\alpha$ against particle
number $N$ for boson stars with $2\omega+3=1/(2k\log\Phi)$ and in
the limit $\Lambda\ra\infty$. All curves are for $k=3$ and are
labelled by the value of $(\si-1)$. The solid portion of each curve
represents stable solutions.}
\end{center} 
\end{figure} 


\begin{thebibliography}{99}

\bibitem{DEF1} T. Damour and G. Esposito-Farese, Class. Quantum
Grav. {\bf 9}, 2093 (1992).

\bibitem{Will2} C. M. Will,
{\it The Confrontation Between General Relativity and Experiment: 
A 1998 Update}, gr-qc/9811036.

\bibitem{Eardley} D. Eardley, Astrophys. J. {\bf 196}, L59 (1975)

\bibitem{Will} C. M. Will, 
{\it Theory and Experiment in Gravitational Physics}, 
Cambridge University Press, 1993.

\bibitem{DamourEF} T. Damour and G. Esposito-Farese, Phys. Rev. Lett
{\bf 70}, 2220 (1993).

\bibitem{DEF2} T. Damour and G. Esposito-Farese, Phys. Rev. D {\bf 54},
1474 (1996).

\bibitem{SalSud} M. Salgado, D. Sudarsky and U. Nucamendi, Phys.
Rev. D {\bf 57}, 124003 (1998).

\bibitem{Zag} H. W. Zaglauer, Astrophys. J. {\bf 393}, 685 (1992).

\bibitem{Gund} M. A. Gunderson and L. G. Jensen, Phys. Rev. D {\bf 48},
5628 (1993).

\bibitem{Torres1} D. F. Torres, Phys. Rev. D {\bf 56}, 3478 (1997).

\bibitem{Comer} G. L. Comer and H. Shinkai, Class. Quantum Grav.
{\bf 15} 669 (1998).

\bibitem{AW} A. W. Whinnett, Class. Quantum Grav. {\bf 16}, 2796
(1999).

\bibitem{AWDT} A. W. Whinnett and D. F. Torres, Phys. Rev. D
{\bf 60}, 104050 (1999).

\bibitem{TSL} D. F. Torres, F. E. Schunck and A. R. Liddle,
Class. Quantum Grav. {\bf 15}, 3701 (1998).

\bibitem{BAT} O. G. Benvenuto, L. G. Althaus and D. F. Torres,
Mon. Not. Roy. Astro. Soc. {\bf 305}, 905 (1999).

\bibitem{FLP1} R. Freiberg, T. D. Lee and Y. Pang, Phys.
Rev. D {\bf 35}, 3640 (1987).

\bibitem{reviews1} P. Jetzer, Phys. Rep. {\bf 220}, 163 (1992).

\end{thebibliography}
\end{document}